\documentclass[preprint,12pt]{elsarticle}
\usepackage{amsmath}
\usepackage{amssymb}
\usepackage{epsfig}
\usepackage{amsmath}
\usepackage{braket}
\usepackage{geometry}
\usepackage{float}
\usepackage{subfig}
\usepackage{graphicx}
\usepackage{mwe}
\usepackage{lipsum}
\usepackage{mathtools}
\usepackage{cuted}

\newcounter{bla}

\journal{Computer Physics Communications}

\begin{document}

\begin{frontmatter}




\title{Simulation of Quantum Many-Body Systems on Amazon Cloud}

\author[a]{Justin A. Reyes}
\author[b]{Dan C. Marinescu}
\author[a]{Eduardo R. Mucciolo\corref{author}}

\cortext[author] {Corresponding author.\\\textit{E-mail address:} mucciolo@physics.ucf.edu}
\address[a]{Department of Physics, University of Central Florida, Orlando, FL 32816, USA}
\address[b]{Department of Computer Science, University of Central Florida, Orlando, FL 32816, USA}

\begin{abstract}
Quantum many-body systems (QMBs) are some of the most challenging
physical systems to simulate numerically. Methods involving approximations for tensor network (TN) contractions have proven to be viable alternatives to algorithms such as quantum Monte Carlo or simulated annealing. However, these methods are cumbersome, difficult to implement, and often have significant limitations in their accuracy and efficiency when considering systems in more than one dimension. In this paper, we explore the \textit{exact} computation of TN contractions on two-dimensional geometries and present a heuristic improvement of TN contraction that reduces the computing time, the amount of memory, and the communication time. We run our algorithm for the Ising model using memory optimized \texttt{x1.32x large} instances on Amazon Web Services (AWS) Elastic Compute Cloud (EC2). Our results show that cloud computing is a viable alternative to supercomputers for this class of scientific applications.
\end{abstract}

\begin{keyword}
tensor network; quantum many-body; cloud computing
\end{keyword}

\end{frontmatter}

\section{Introduction}
\label{Introduction}

Quantum many-body (QMB) physics is concerned with the study of
microscopic systems involving a large number of interacting particles
\cite{Bruus2002}. Studies of QMB have applicability across a wide range
of physics, chemistry, and material science problems~\cite{Baumgartner2005, Murg2010, Ghosh2014}, such as the study of magnetism, superconductivity, topological order, and spin glasses~\cite{Auer2005}.

Several tensor network (TN) algorithms have been formulated to implement
computer simulation of these systems~\cite{Auer2005,Verstraete2004,Baumgartner2002a,Baumgartner2002b,Pfeifer2010,Verstraete2008,Vidal2008}. For the one dimensional case, simulations can scale up to large system sizes because TN contractions can be carried out exactly with a polynomial effort, but this is not the case for systems of higher dimension. Simulations for two- and three-dimensional systems have been limited to particular geometries and systems of modest size due to the overwhelming number of computational resources required by tensor contractions in more than one dimension~\cite{Lam1997,Schuch2007}. 

Approximate methods to overcome these limitations have been proposed, for instance, by simplifying the tensor environment, thus, avoiding a full contraction~\cite{Orus2009,Gu2008,Levin2007,Xie2009}. However, often the focus of the literature has been on infinite, translation-invariant systems and no particular attention has been paid to adapt the methodology to distributed computing. In this paper, we investigate two-dimensional QMB spin systems and explore the limitations of a cloud computing environment using on a heuristic for parallel TN contractions without approximations. For this study we use the Amazon Web Services (AWS) Elastic Compute Cloud (EC2) instances.
  
Tensors are algebraic structures represented as
multi-dimensional arrays. The order of a tensor is defined by the
number of indices needed to specify a tensor element. For
instance, scalars are tensor of zero order, while vectors and matrices
are tensors of order one and two, respectively. A tensor network (TN)
is a decomposition of a high-order tensor into a set of low-order
tensors which share indices under a specified geometry. To extract
information from a TN, it is necessary to perform a summation over all
shared indices, a procedure termed {\it tensor contraction}.

TNs are now ubiquitous in QMB simulations because they provide a
systematic way to represent and approximate quantum wave
functions. These wave functions can have exponentially many
components. For instance, the Hilbert space of a QMB system of $N$
interacting spin-1/2 particles (each being a two-state subsystem) has
dimension $2^N$ on the spin sector. A tensor representation for such a
system requires a tensor of order $N$ where each index represents a
spin in the lattice. A more compact and often more efficient
representation is obtained by decomposing this high-order tensor into a
TN, where shared tensor indices are representative of the entanglement
between particles generated by their interactions.

Even though the Hilbert space of a QMB system grows exponentially with
the number of degrees of freedom, it is also often the case that only
a small region of that space contains useful information about the
state of the system \cite{Poulin2011}. This is particularly the case for
systems containing local interactions and whose ground-state energy
has a gap separating it from excited states; this gap does not
scale with system size. Thus, to find the ground state of such gapped
systems involving local interactions, only a small region of the
Hilbert space needs to be included in the system's ground state wave
function. This phenomenon is captured by how the amount of
entanglement between one subsystem and another scales with their
sizes. If the entanglement entropy scales only with the size of the
boundary separating the two subsystems, the so-called ``area law''
\cite{Eisert2010}, an efficient TN representation of the ground state is
possible, with tensors being polynomially bounded. In two-dimensional cases, particularly near critical points of QMB systems, the entanglement entropy scales with a logarithmic correction to the boundary of the smallest subsystem. In these cases, even with only a logarithmic correction to the area law, high-dimensional TN computations are known to require an exponential effort because of the exponential growth of internal bond dimensions during TN contraction.

To advance the study of QMB systems, even away from critical points,
it is necessary to optimize TN contractions. The optimization of TN
contractions is a currently extensive area of study
\cite{Ghosh2014,Baker1994,Khuller2002,Leighton1999,Seymour1994,Ibrahim2017}. Of particular importance, it is noted that the order of
summations taken is critical to the determination of the computational
time necessary for a contraction. Orus \cite{Orus2009} demonstrates the importance of this by comparing two different orderings for the contraction of three tensors, each with indices taking $D$ values. Finding an optimal order is NP-hard problem \cite{Markov2008}. Equally important,
as we will show in this paper, is the partitioning of the system into
various independent parallel contractions.

The creation of libraries such as the Cyclops Tensor
Framework (CTF) \cite{Solomonik2012} and the development of  Tensor Processing Units (TPU) are advantageous for handling tensor-tensor operations, but have yet to address some of the major concerns for QMB simulations using TN contractions on compute clouds. CTF partitions large tensors cyclically and would allow computation of larger problems, but it also would significantly increase the number of messages exchanged among concurrent threads. This is an undesirable side-effect for computations on systems with large communication latency. Moreover; partitioning of individual tensors across multiple processors in CTF is too general, failing to take advantage of the geometry of a physical system

TPUs are optimized for matrix-matrix operations and speed up computations, but do not accommodate the exponential growth of tensors throughout the contraction. What is needed for efficient contraction without approximation is a set of processors with a large cache and physical memory and a partitioning of the TN across processors in a systematic and geometrically advantageous way. We focus our attention on implementing parallel geometric TN partitioning on AWS EC2 instances with large memories.

This paper is organized as follows: In Sec. \ref{QMBSimulation}, we
discuss progresses in cloud services that have motivated their use for
the simulation of QMB systems. In Sec. \ref{sec:IsingModel}, we
describe the QMB system model used for the numerical calculations. In
Sec. \ref{TNC}, we examine the computational difficulties involved in
the contraction of a two-dimensional TN representing a many-body spin lattice system and present a heuristic for parallel contraction. In Sec.
\ref{QMBComputations}, we compare various algorithms for the
contraction of TNs of the spin system, and in
Sec. \ref{ExperimentalResults} we present the performance analysis of
our algorithm on the specific EC2 x1e instance type, along with some
benchmark results for the spin system. The conclusions are given in
Sec. \ref{Conclusions}.
\section{Cloud QMB Simulation} 
\label{QMBSimulation}

The cloud computing infrastructure is designed to perform optimally
for Big Data, online transaction processing, and data streaming
applications. Such applications exploit data-level, task-level, and
thread-level parallelism. The Warehouse scale computers (WSCs), which
serve as the backbone of the cloud infrastructure, host tens to
hundreds of thousands of servers communicating through networks with
sufficient bandwidth but also with a relatively high latency. This
infrastructure is advantageous for the enterprise cloud applications, where procedures such as serialization, deserialization, and the compression of buffers, and Remote
Procedure Calls (RPCs) account for only 22--27\% of the CPU cycles. For these applications, this cost is a nominal ``WSC architecture tax`` \cite{Kanev2015, Marinescu2017}

However, for scientific and engineering applications, the communication latency can produce a significant effect on the performance. QMB applications can exhibit a fine grained parallelism, deploying many parallel threads communicating frequently with each other, and using barrier synchronization to transit from one stage of computation to the next. While there are methods such as the parallel Monte-Carlo which have been optimized for such communication costs~\cite{Kent2003}, many other methods simply cannot avoid this level of communication. This is particularly the case for parallel TN algorithms, such as the parallel-DMRG~\cite{Stoudenmire2013}, where partial computations must be merged frequently. Because of the communication latency of current cloud services, QMB applications are optimally run on
supercomputers with fast interconnection networks such as Mirinet,
Infiniband, or some custom designed network. For instance, a group from
ETH Zurich was able to perform a 45-qubit simulation using a
supercomputer at the Lawrence Berkeley National Laboratory
\cite{Haner2017}. 

In recent years, Cloud Service Providers (CSPs) have managed to narrow the
performance gap vis-a-vis supercomputers. Clusters with faster
interconnects and instances with physical memory on the
order of hundreds of GiB are now available. Also, faster and more powerful processors and coprocessors are being developed for new instance types. For example, Amazon provides Graphics Processing Units (GPUs), optimal for linear algebra operations, while Google has pioneered the Domain Specific
Architectures (DSAs) with the creation and provision of Tensor
Processing Units (TPUs), which are optimal for the many small
matrix-matrix operations needed in deep learning
applications. However, the challenges of managing high order tensor
contractions in QMB applications are still present. Both the
number of operations and the memory footprint involved in the
computation grow exponentially with the system size. 

For QMB simulations, we choose \texttt{x1.32x large} EC2 instances with the largest caches and storage space. This strategy reduces the amount of communication among parallel threads carrying out tensor contractions concurrently. We avoid distributing a single tensor across multiple threads, choosing instead to distribute groups of
tensors according to the geometry of the system considered.

\section{The QMB model system}
\label{sec:IsingModel}

We adopted the spin-1/2 Ising model in
the presense of a transverse field for simulation of the QMB system. The Ising model is used in statistical mechanics to describe magnetic lattice systems with strong anisotropy \cite{Stinchcombe1973}. Recently, it has been used as a paradigm for the study of quantum phase transitions
\cite{Sachdev2011}. The model consists of discrete variables $S_i^z$
that represent magnetic dipole moments of atomic spins that can be in
one of two states $(+1)$ or $(-1)$. The spins are arranged in a
lattice, allowing each spin to interact with its nearest neighbors
(spin-spin interaction). For spatial dimensions larger than one, the
model has a finite-temperature phase transition and critical
behavior.

When a transverse magnetic field is present, the model yields a
zero-temperature phase transition (i.e., a quantum phase transition)
driven by the competition between the spin-spin interaction, which
favors ferromagnetism (if $J>0$) or antiferromagnetism (if $J<0$), and
the external field, which favors paramagnetism. In this paper, we
consider the case where the spins are located on a rectangular
lattice. Mathematically, the model is defined by the Hamiltonian
(total energy) of the system,
\begin{equation}
\label{eq:Hamiltonian}
\hat{H} = \hat{H}_J + \hat{H}_{\Gamma},
\end{equation}
with the two terms
\begin{equation}
\hat{H}_J = -J \sum_{\langle i,j \rangle} \hat{S}_i^z \hat{S}_j^z
\end{equation}
and 
\begin{equation}
\hat{H}_{\Gamma} = -\Gamma \sum_i \hat{S}_i^x
\end{equation}
describing, respectively, the spin-spin interactions between
nearest-neighbor sites $\langle i,j \rangle$ and the coupling of the spins to a transverse field (here denoted by $\Gamma$). The constant $J$ quantifies the spin-spin interaction. Different components of the on-site spin operator do not commute, namely, $\hat{S}_i^z\, \hat{S}_i^x \neq
\hat{S}_i^x\, \hat{S}_i^z$, lending the two terms in the Hamiltonian
non commuting. It is this noncommutability that yields quantum
critical behavior and entangled many-body states for the spin
system. At zero temperature, one finds a critical point when $\Gamma
\approx 3J$ \cite{duCroodeJongh1998}. At this point, the spins in the
system are highly entangled.

The ground state energy $E_0$ corresponds to the lowest eigenvalue of
the operation $\hat{H}$. All other eigenvalues are associated to
excited states.

The vector describing the wave function of the system can be written
as
\begin{equation}
\label{eq:Psi}
  \ket{\Psi} = \sum_{\lbrace \sigma_k \rbrace} A(\lbrace \sigma_k
\rbrace) \ket{\sigma_1 \cdots \sigma_N},
\end{equation}
where $\sigma_k=\pm 1$, with $k=1,\ldots,N$ indicating the $N$ spin
degrees of freedom. The connection between these variables and
conventional binary ones is straightforward: $x_k =
(\sigma_k+1)/2$. Notice that there are $2^N$ basis vectors
$\ket{\sigma_1 \cdots \sigma_N}$. The amplitudes $A(\lbrace \sigma_k
\rbrace)$ are in general complex numbers; however, for the model in
consideration they can always be defined as real. The eigenvector
$|\Psi_0\rangle$ associate to $E_0$ yields the ground state wave
function of the system, namely, $\hat{H} | \Psi_0 \rangle = E_0
|\Psi_0 \rangle$. (In the absence of the external field, the ground
state is two-fold degenerate due to the spin inversion symmetry of
$\hat{H}_J$.)

An important quantity associated with the state vector is its norm,
\begin{equation}
  \label{eq:norm}
\langle \Psi | \Psi \rangle = \sum_{\lbrace
  \sigma_k \rbrace} A^\ast(\lbrace \sigma_k \rbrace)\, A(\lbrace
\sigma_k \rbrace).
\end{equation}
There are also a number of physical quantities of importance that can
be obtained from the state vector. The expectation value of the total
energy of the system is defined as
\begin{equation}
  \label{eq:Etot}
E = \frac{\langle \Psi | \hat{H} | \Psi \rangle} {\langle \Psi | \Psi
  \rangle}.
\end{equation}
The local transverse and longitudinal magnetizations of the system at
site $k$ are given by
\begin{equation}
  \label{eq:mx}
m^x_k = \frac{\langle \Psi | \hat{S}^x_k | \Psi \rangle} {\langle \Psi
  | \Psi \rangle}
\end{equation}
and
\begin{equation}
  \label{eq:mz}
m^z_k = \frac{\langle \Psi | \hat{S}^z_k | \Psi \rangle} {\langle \Psi
  | \Psi \rangle},
\end{equation}
respectively. Finally, the longitudinal spin-spin correlation between
spins at sites $i$ and $j$ is equal to
\begin{equation}
  \label{eq:cij}
c_{ij} = \frac{\langle \Psi | \hat{S}^z_i\, \hat{S}^z_j\, | \Psi
  \rangle} {\langle \Psi | \Psi \rangle} - \frac{\langle \Psi |
  \hat{S}^z_i | \Psi \rangle \langle \Psi |
  \hat{S}^z_j\, | \Psi \rangle} {\langle \Psi | \Psi
  \rangle^2}.
\end{equation}
By representing the amplitudes $A(\{\sigma_k\})$ as a tensor network,
all the physical quantities above can be computed via suitable tensor
network contractions.

\section{Tensor Network Contraction}
\label{TNC}

\begin{figure*}[htb!]
\begin{center}
\includegraphics[width=12cm]{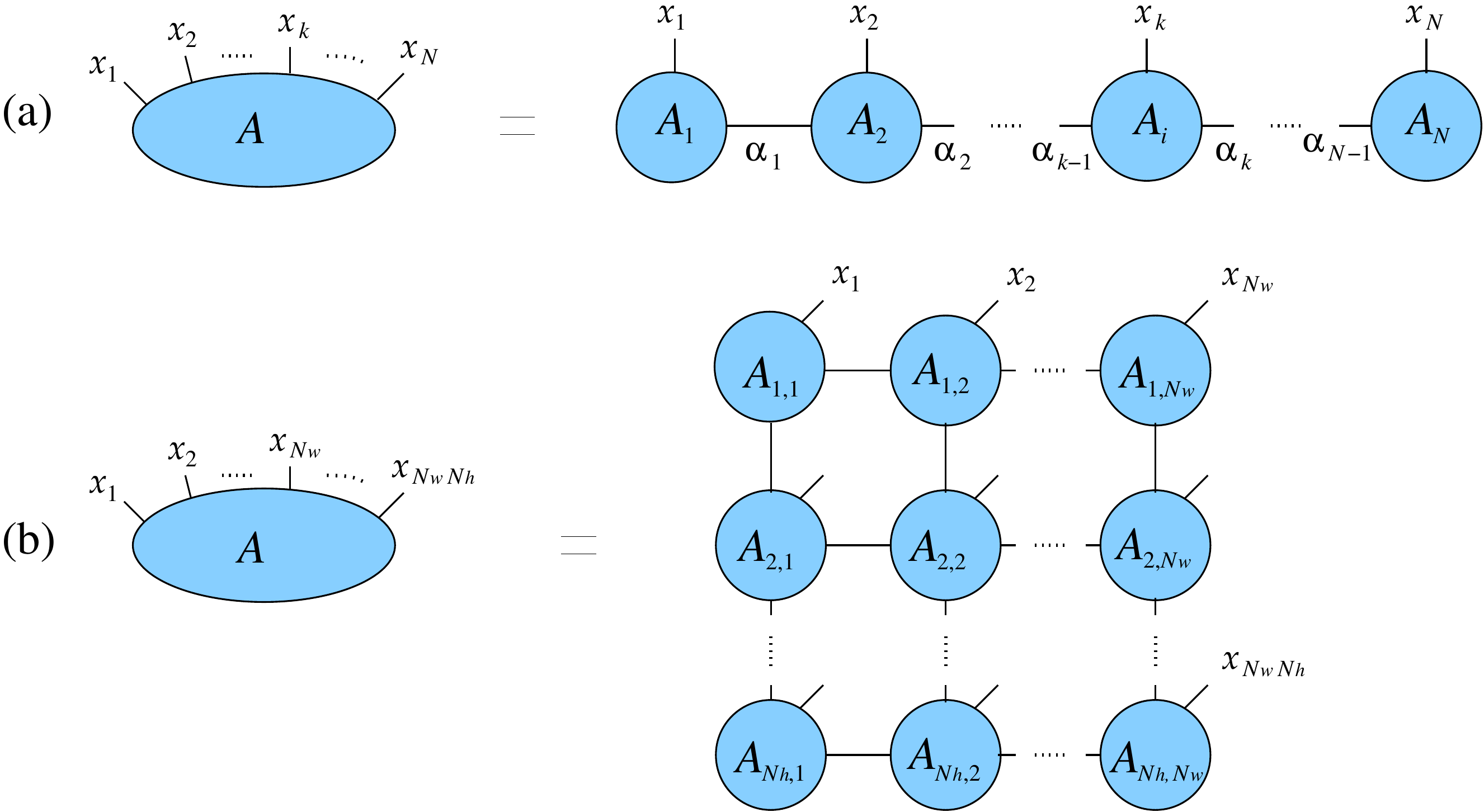}
\end{center}
\caption{(a) The matrix-product state decomposition of a tensor $A$ of
  order $N$. (b) The rectangular lattice decomposition of a tensor $A$
  of order $N = N_h\, N_w$.}
\label{fig:TensorDecomp}
\end{figure*}

As previously mentioned, a TN is a decomposition of a high-order
tensor into a set of low-order tensors sharing internal indices under
a specific geometry. Consider, for instance, a tensor with index set
$\{x_k\}_{k=1,\ldots,N}$, with $x_k=0,1$ for all $k=1,\ldots,N$, whose
elements are expressed as $ \left[ A \right]
_{x_1,x_2,\ldots,x_N}$. This tensor has $2^N$ elements thus require an
exponential amount of storaga memtory. A possible decomposition of
this tensor is given by the expression
\begin{equation}
\label{eq:Adecomp}
 \left[ A \right] _{x_1,x_2 ... x_N} = \sum_{\alpha_1,\ldots,\alpha_N}
 \left[ A_1 \right] _{x_1}^{\alpha_1} \left[ A_2 \right]
 _{x_2}^{\alpha_1 \alpha_2} \cdots \left[ A_N \right] _{x_N}^{\alpha_{N-1}
   \alpha_N},
\end{equation}
where $\{\alpha_k\}_{k=1,\ldots,N}$, are the internal (repeated)
indices of the network, with each index $\alpha_k=1,\ldots,\chi_i$ for
a suitable $\chi_k$ (often referred as bond dimension). This
particular chain decomposition is known as a matrix product state
(MPS) \cite{Perez-Garcia2006}, see Figure \ref{fig:TensorDecomp}a. An MPS
is obtained by repeated applications of singular-value decomposition
operations on the original tensor $A$. Notice that each index $x_i$
now resides on an individual tensor $A_i$ of order two (for $k=1,N$)
or three ($k=2,\ldots,N-1$). If $\chi_k \sim {\rm poly}(N)$, Equation
(\ref{eq:Adecomp}) provides a compact decomposition of tensor $A$,
requiring only a polynomial amount of storage space.

A matrix product state is not the only possible decomposition of a
tensor. Consider, for instance, the case when $N = N_h\, N_w$. One can
then decompose the tensor $A$ into a $N_w \times N_h$ rectangular
lattice, see Figure \ref{fig:TensorDecomp}b. As mentioned in
Sec. \ref{Introduction}, for representative classes of QMB systems, it
is indeed the case that bond dimensions are polynomially bounded, and
the tensor network decomposition of a ground-state wave function
provides a compact representation when performed appropriately. The
most suitable decomposition minimizes the bond dimensions $\chi_k$ and
is determined by the interactions present and the system geometry. For
instance, the decomposition in Figure \ref{fig:TensorDecomp}b is
particularly useful for the representation of the quantum amplitude
$A(\{\sigma_k\})$ of the Ising model wave function, see
Eq. (\ref{eq:Psi}), as each tensor in the lattice can be associated to
one physical spin.

To contract a tensor decomposed into a network, it is necessary to
perform a summation over all the internal (repeated) indices in the
network. Tensor contraction can be done in different ways depending 
upon a number of factors including the network topology.

By virtue of our choice of QMB system, in this paper we consider only
planar, rectangular TNs, as shown in the example of
Figure \ref{fig:TensorDecomp}b. In addition, we focus on the computation
of scalar quantities such as those defined in Eqs. (\ref{eq:norm}) to
(\ref{eq:cij}), which can be cast as the contraction of two planar TNs
(one for $A^\ast$ and another for $A$) into a single planar TN with no
external indices. Thus, the computation of physical quantities
requires only the sum over all internal indices of a planar TN.

More specifically, consider a square lattice of tensors with size
$L\times L$. The tensors on the four corners of the lattice are of
order two, those along the edge are of order three, and those within
the bulk are of order four. In practice, the full contraction of a TN
occurs by contracting tensors pairwise and sequentially. The number of
tensor elements (we call dimension), an important quantity for
determining memory requirements, evolve as follows. Consider the
contraction of two tensors, $A_1$ and $A_2$, with dimensions $d(A_1)$
and $d(A_2)$, respectively, into a tensor $B$. The dimension $d(B)$ of
the resulting tensor satisfies
\begin{equation}
d(B) = \frac{d(A_1)*d(A_2)}{d({x})} ,
\end{equation} 
where $x$ is the set of shared indices between $A_1$ and $A_2$. Thus,
the tensor dimension can increase substantially after a pairwise
contraction. For every full TN contraction there exists a {\it
  bottleneck contraction}, after which every tensor pair contraction
no longer increases the memory footprint. It is our desire to minimize
the size of this bottleneck, which in turn optimizes the memory
requirements and the number of floating point operations (FLOPs)
necessary to perform the computation.

\begin{figure}[htb!]
\begin{center}
\subfloat[]{\includegraphics[scale=0.4]{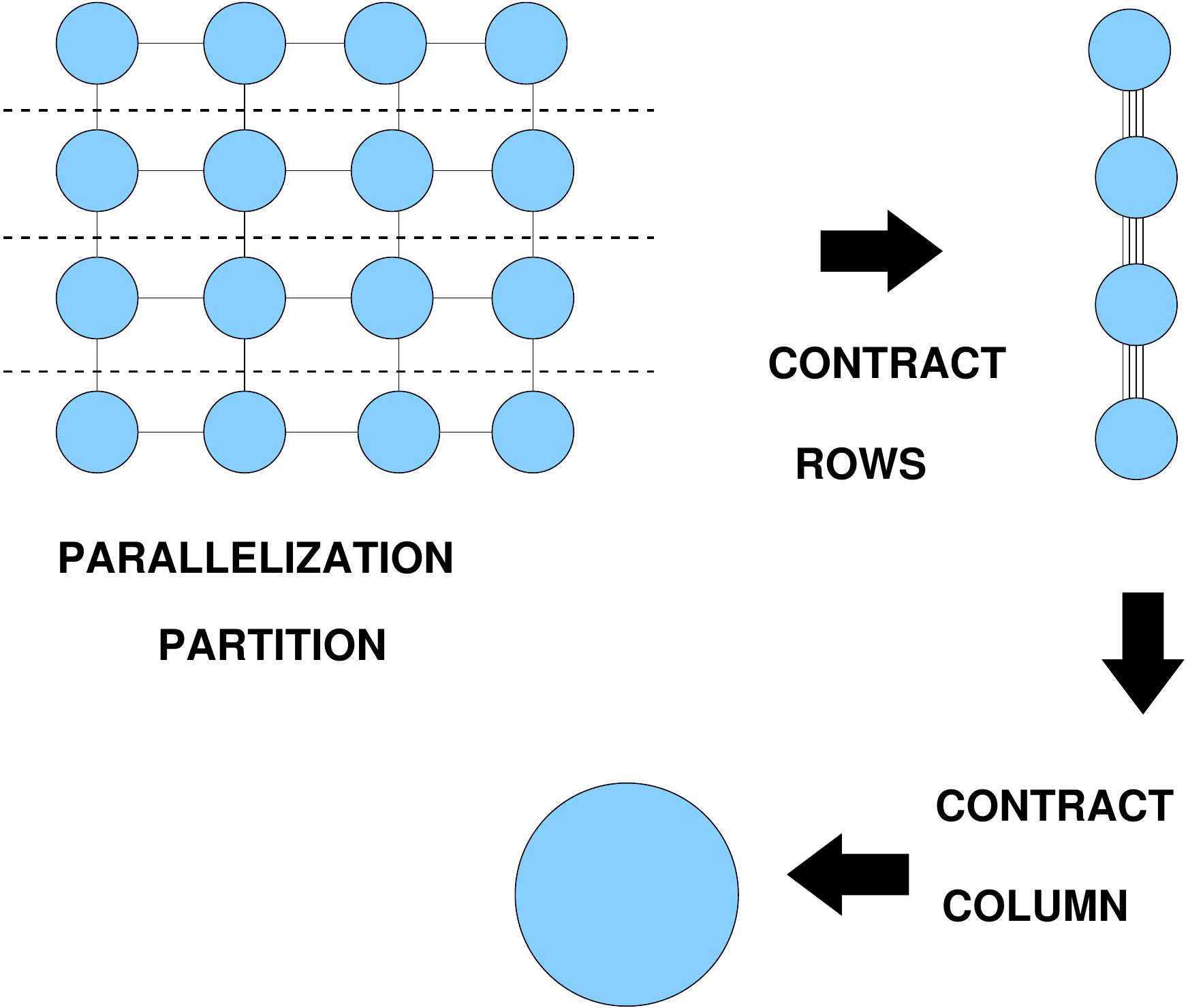}}\\
\subfloat[]{\includegraphics[scale=0.4]{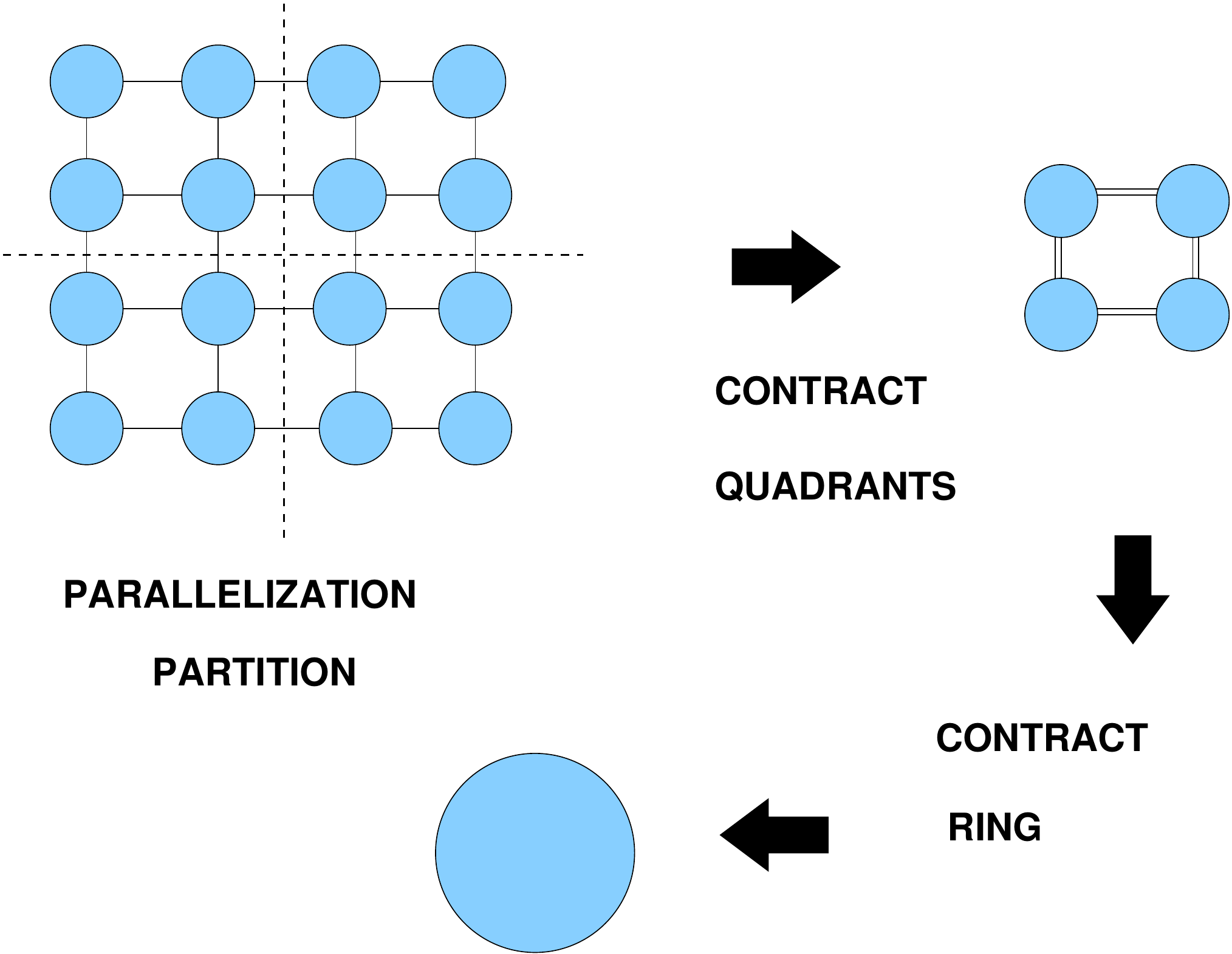}}
\end{center}
\caption{A graphical representation of the (a) row and (b) quadrant  contraction algorithms for a square lattice with uniform bond dimension. The number of messages sent in the row scheme is dependent on the number of rows, while the number of messages in the quadrant scheme remains constant, with only four messages being passed between concurrent processes.}
\label{QuadrantCompression}
\end{figure}

Naively, we could consider simply contracting the square lattice along
the rows from one edge to the other. Assuming the dimension of each
initial index was $\chi$, this would leave a chain of tensors each
having $\chi^{L}$ elements, which could then be contracted as an
MPS. However, this is not the only option. If instead we define four
quadrants for the lattice and contract from each edge to the midpoint
of the lattice, we end with a ring of tensors each having $\chi^{L/2}$
elements, which can then be fully contracted as an MPS. Another option
yet is to contract from the edge somewhere between these two previous
approaches, ending up with at least one tensor in the final ring with
more than $\chi^{L/2}$ elements. Thus, as mentioned earlier, the order
of the contraction matters.

In addition to selecting an appropriate contraction order, we
partition the lattice for parallel computation according to both the
geometry of the lattice and contraction order. In Figure ~\ref{QuadrantCompression}, this is demonstrated for both of the previously
discussed contraction orders on a square lattice with linear size $L=4$. Notice that the partitioning of the quadrant scheme allows for fewer number of messages between processes. The number of messages sent in the row scheme is dependent on the number of rows, while the number of messages in the quadrant scheme remains constant, with only four messages being passed between processes. 

Our choice of contraction ordering and geometry-specific partitioning
can be extended to multiple geometries. The heuristic is as follows:
given a TN geometry, select the order which minimizes the bottleneck
contraction size and partition the lattice around the ring of tensors
for parallel computation with minimal communication. In
Figure \ref{ExecutionTimeQCvsRC}, we compare these parallel contractions to the contraction of square lattice where individual tensors are
cyclically parallelized, in contrast to our parallelization of groups
of tensors within the lattice.

\section{QMB Ground State Computation}
\label{QMBComputations}

In this section we discuss a few of the fundamental concepts behind
the determination of the ground state of a QMB systems using TNs.

A particularly important family of TN states are projected entangled
pair states (PEPS) \cite{Verstraete2008}. In contrast to the
one-dimensional nature of MPS, PEPS correspond to higher dimensional
tensor networks, e.g., two-dimensional TNs~\cite{Verstraete2004,Orus2009,Levin2007,Xie2009,Jordan2008,Murg2007}. For gapped systems with local
interactions, PEPS is particulaly suitable since: (a) {\it bond
  dimensions} of tensors in the TN -- the range $\chi$ of which the
indexes run -- can be kept small and yet provide an accurate
description of low-energy states; PEPS naturally satisfy the area-law
scaling of the entanglement entropy; and (d) PEPS can handle
reasonably well two-point correlation functions that decay
polynomially with the separation distance~\cite{Murg2010}.

Following the scheme shown in Figure \ref{fig:TensorDecomp}b, for our
simulations we associate to each spin in the bulk of the
two-dimensional lattice a tensor of rank 5. Four indices of this
tensor account for the bonds to nearest-neighbor spins. The fifth
index accounts for the binary nature of the spin variable. For spins
located at the edges and corners of the lattice, only three and two
bonds are required, respectively. When all bond (internal) indices are
contracted, the resulting scalar quantity yields the probability
amplitude $A(\lbrace \sigma_k \rbrace )$ of finding the spin system in
the particular basis state $\lbrace \sigma_k \rbrace$, $k = 1,...,N,$,
see Eq. (\ref{eq:Psi}).

A standard algorithm for the determination of the ground state energy
of a PEPS is the imaginary time evolution (ITE) algorithm. This
algorithm is defined by the iterative application of an incremental
imaginary time evolution operator $\hat{U}_{\delta \tau} =
e^{-i\hat{H}\,(-i\delta \tau)} = e^{-\hat{H}\,\delta \tau}$ to an
initial quantum state $\ket{\Psi_{\rm initial}}$, over $m$ steps. This
process evolves the system's state in imaginary time in incremental
steps $\delta \tau = \tau/m$, where $\tau$ is the total time of the
evolution. For $\tau \gg \mbox{max}\{J^{-1},\Gamma^{-1}\}$, the state
vector $|\Psi\rangle = \left[ \prod_{i=1}^m \hat{U}_{\delta\tau}
  \right] |\Psi_{\rm initial}\rangle$ becomes exponentially close to
the ground state $|\Psi_0\rangle$, provided that the initial state had
a nonzero overlap with the ground state, namely, $\langle \Psi_{\rm
  initial} | \Psi_0\rangle \neq 0$. Typically, one chooses as the
initial state $|\Psi_{\rm initial}\rangle$ a product state that is a
random superposition of individual spin states. We initialize our TN
to a uniform superposition of product states,
\begin{equation}
|\Psi_{\rm initial}\rangle = \frac{1}{2^{N/2}} \prod_{k=1}^N \left[
  \sum_{\sigma_k=\pm 1} |\sigma_k \rangle \right]
\end{equation}
(i.e., the uniform bond dimension is initially $\chi = 1$).

Each iteration of the time evolution can be divided into subintervals
defined by the application of noncommuting operators in the
Hamiltonian $\hat{H}$. For our example, the Hamiltonian $\hat{H}$ in
Eq. (\ref{eq:Hamiltonian}) is a summation of the noncommuting
operators $\hat{H}_J$ and $\hat{H}_{\Gamma}$. If we consider the
second-order Trotter-Suzuki approximation \cite{Suzuki1976} of
$\hat{U}_{\delta\tau}$, then we obtain
\begin{equation}
\hat{U}_{\delta\tau} = \hat{U}^{\Gamma}_{\delta\tau/2}\;
\hat{U}^{J}_{\delta\tau}\; \hat{U}^{\Gamma}_{\delta\tau/2},
\end{equation} 
where $\hat{U}^{\Gamma}_{\delta\tau/2} = e^{-\hat{H}_{\Gamma}
  (\delta\tau/2)}$ and $\hat{U}^{J}_{\delta\tau} = e^{- \hat{H}_{J}
  \delta\tau}$. These operators naturally define three subintervals
according to the operators involved within each time step
$\delta\tau$.

To accurately monitor the convergence of the algorithm to the optimal
state, the expectation value of the total energy of the system is
periodically evaluated every two $\delta\tau$ steps, see
Eq. (\ref{eq:Etot}). For the Ising model, this results in the
calculation of the local magnetization $m_i^z$, see Eq. (\ref{eq:mz}),
and the and two-spin correlation $c_{ij}$, see Eq. (\ref{eq:cij}), for
each spin site and site pair, respectively. This too defines two
additional sub-steps during the algorithm.

After applying the sets of operators
$\lbrace\hat{U}^{\Gamma}_{\delta\tau/2}\rbrace$ to every site or
$\lbrace \hat{U}^{J}_{\delta\tau}\rbrace$ to every pair, the system is
normalized, $|\Psi\rangle \rightarrow |\Psi\rangle/\sqrt{\cal N}$,
maintaining both the stability of the algorithm and the probabilistic
interpretation of $\ket{\Psi}$.

Within a single time step, the actual order of events is as follows:
a) update the system by applying the one-body operators to every site,
and then normalize. b) update the system by applying the two-body
operators to every pair of sites, and then normalize. c) after every
two iterations of applying operators to the system, calculate the
expectation value of the total energy of the system. With all of this
in mind, it is now understood that each stage of the imaginary time
evolution algorithm allows for the modification or evaluation of the
system in three distinct ways:
\begin{enumerate}
\item a tensor can be locally updated by the application an operator;
\item the norm of the system can be calculated, or
\item an expectation value can be calculated.
\end{enumerate}
We now examine how each of these three tasks are accomplished.

To update the system and push it towards the ground state, the
operators $\hat{U}^{\Gamma}_{\delta\tau/2}$ and
$\hat{U}^{J}_{\delta\tau}$ are applied iteratively as previously
mentioned. These operators are themselves composed of a series of
local operators applied sequentially to either each single site or to
pair of sites in the lattice. For the one-body operators
$\hat{U}^{\Gamma (i)}_{\delta\tau/2}$ acting on site $i$, a
tensorization of the operator is constructed and then contracted along
the physical index of site $i$. Full contraction of the lattice is
unnecessary at this stage.

\begin{figure}[htb!]
\begin{center}
\includegraphics[scale=0.6]{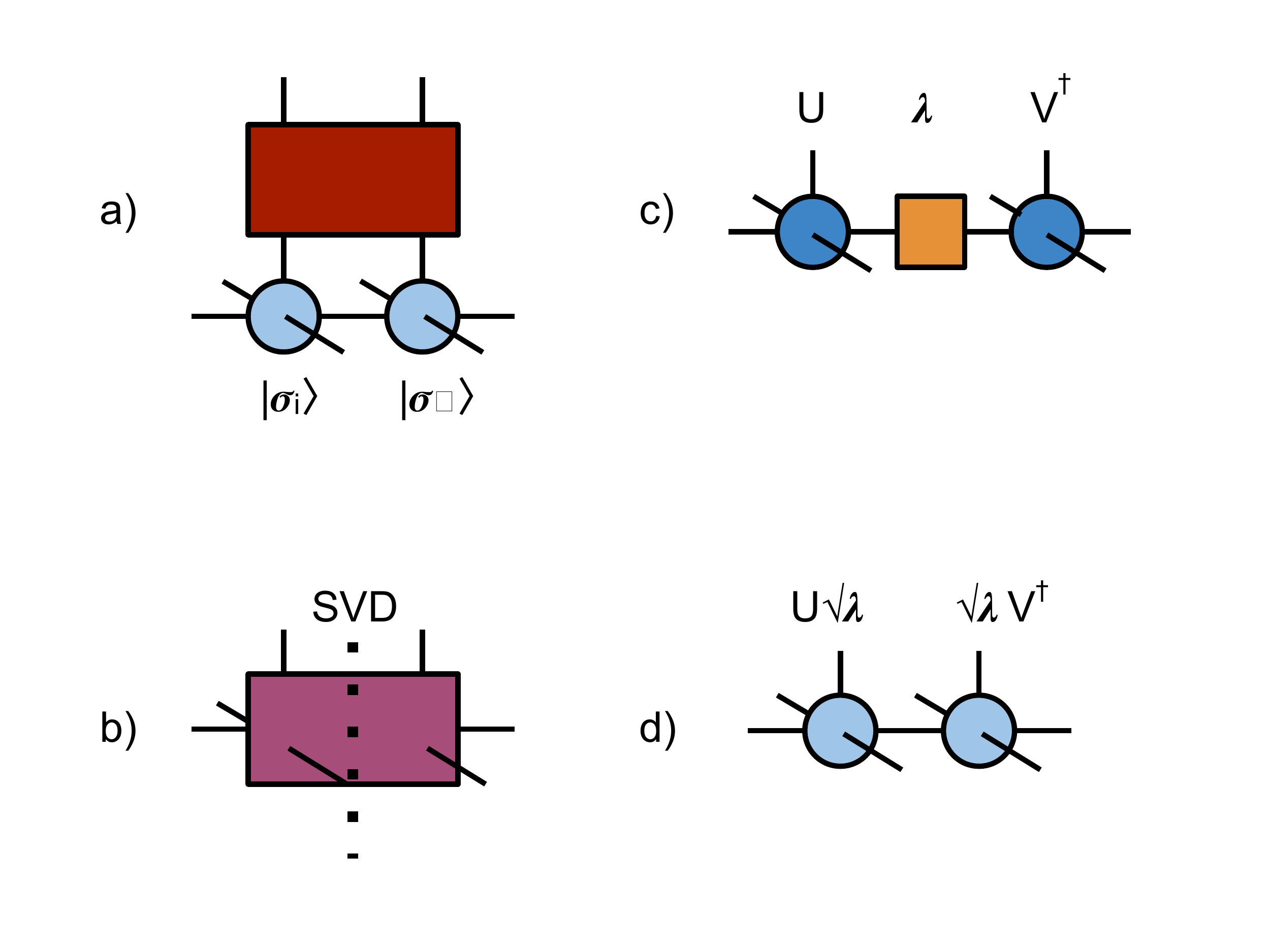}
\end{center}
\caption{A graphical representation of the singular value
  decomposition step iterated as follows: (a) operator
  $\hat{U}^{J(i)}_{\delta\tau}$ (red) is applied to two sites (light
  blue), contracting along the spin indices; (b) the resultant
  tensor is matricized with the remaining open indices and (c) decomposed by singular value decomposition; (d) the
  initial pair of sites is replaced with the matrices $U$ and $V^{\dagger}$ right and left multiplied, respectively with the square root of the singular values $\sqrt{\lambda}$.}
\label{fig:decomposition}
\end{figure}

For the two-body operators $\hat{U}^{J(ij)}_{\delta\tau}$ acting on a
pair of sites $ij$, the update includes an additional step. First, the
tensorization in the tensor product basis of the pair's physical
indices is constructed. Then the contraction over the physical indices
is performed. The resultant tensor is then decomposed back into the
two individual sites in their respective basis by performing a
singular value decomposition, as shown in Figure
\ref{fig:decomposition}. The singular value decomposition is a
necessary step in allowing the exchange of information throughout the
lattice, which, in turn enables an increase in the entanglement
entropy. This increase in the entanglement entropy leads to the growth
of bond dimensions. If the system is far away from the critical point,
i.e., the gap between the ground state and the first excited state is
finite and system-size independent, the imaginary-time evolution
encounters only a low entanglement growth (i.e. small bond dimensions
are maintained as the system size grows). 

\begin{figure}[htb!]
\begin{center}
\includegraphics[scale=0.35]{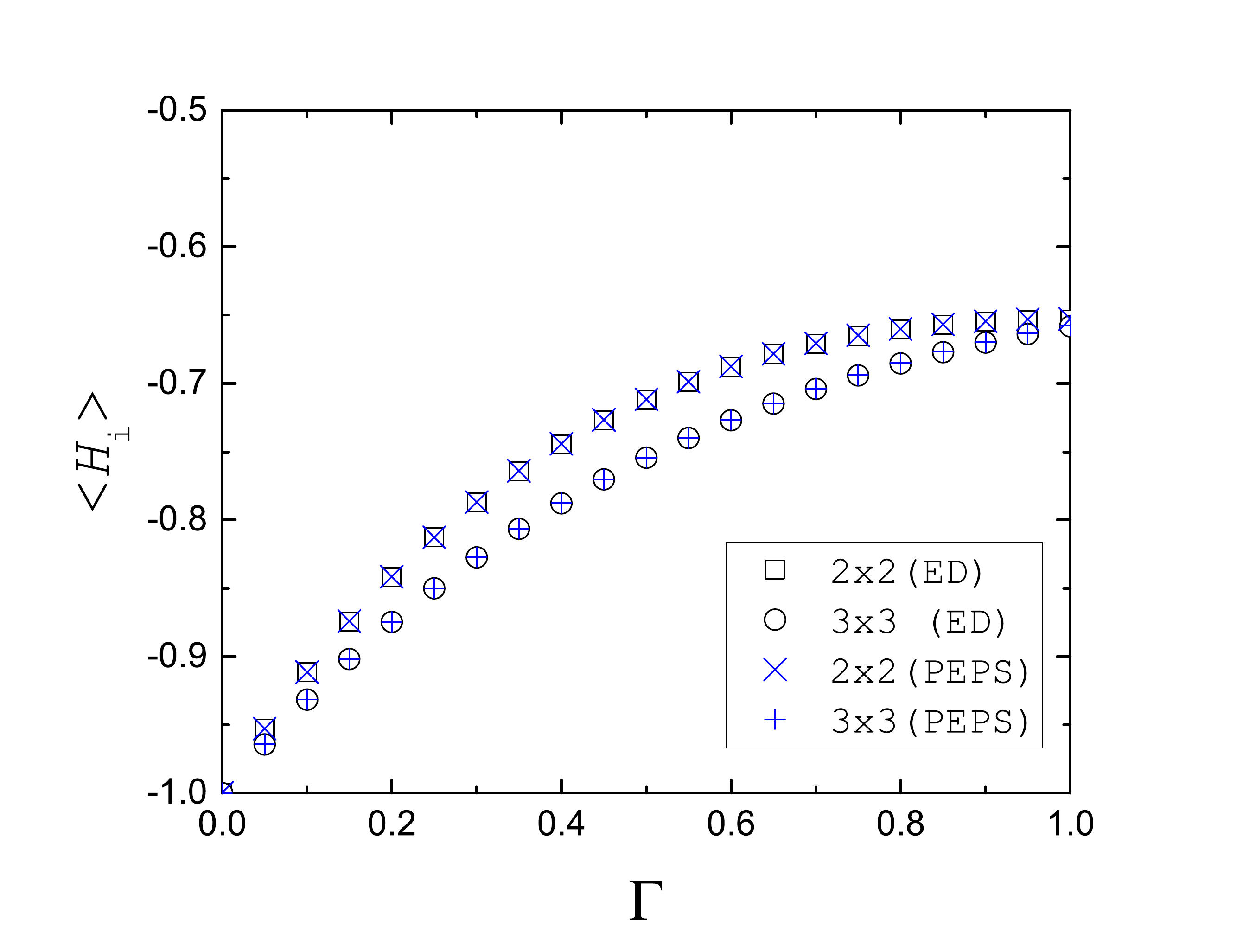}
\end{center}
\caption{A comparison of the average final energy per site calculated
  using the imaginary time evolution algorithm of a tensor network
  (PEPS) versus exact diagonalization of the full Hamiltonian (ED) for
  lattices of size $2\times 2$ and $3\times 3$ with maximal bond
  dimension of $\chi = 2$. The ITE was run for $\tau = 3$ and 200
  steps ($J=1$). It is evident that under these parameters and in this
  regime of transverse field values, the low entanglement
  approximation (i.e. small bond dimensions) is sufficient.}
\label{fig:energy_comparison}
\end{figure}

However, as the critical
point is approached, there is a significant increase in the
entanglement. Entanglement scales up significantly with the system
size in this case, significantly increasing the bond dimension
necessary for accurate simulation. {\it The interplay between bond
  dimension and lattice size is of fundamental importance to the
  determination of what types of problems are practically solvable.}
As mentioned earlier, the quantum critical point for the
two-dimensional Ising model with transverse field occurs where the
quantity $\Gamma/J \approx 3$. Figure ~\ref{fig:energy_comparison} shows that for small lattices far from this point, the final state
obtained while allowing only minimal bond growth is in good agreement
with the solutions obtained from exact diagonalization. We therefore
primarily focus on systems where $\Gamma/J$ is far from the critical
point.

After applying the nonunitary operators
$\hat{U}^{\Gamma}_{\delta\tau/2}$ and $\hat{U}^{J}_{\delta\tau}$ to
the two-dimensional tensor lattice, the state must be renormalized to
keep the algorithm numerically stable. The norm is calculated simply
by contracting the tensor network state $\ket{\Psi}$ with its
conjugate $\bra{\Psi}$ along every spin index. The network must be
similarly fully contracted for every site and pair in the lattice to
calculate $E$, $m_i^x$, and $c_{ij}$. For each lattice site $i$ the
state $\hat{S}^x_i\ket{\Psi}$ is contracted with $\bra{\Psi}$. For
each pair $\langle ij \rangle$, the state $\hat{S}^z_i
\hat{S}^z_j\ket{\Psi}$ is contracted with $\bra{\Psi}$.

\begin{figure}[htb!]
\begin{center}
{\includegraphics[width=0.6\linewidth]{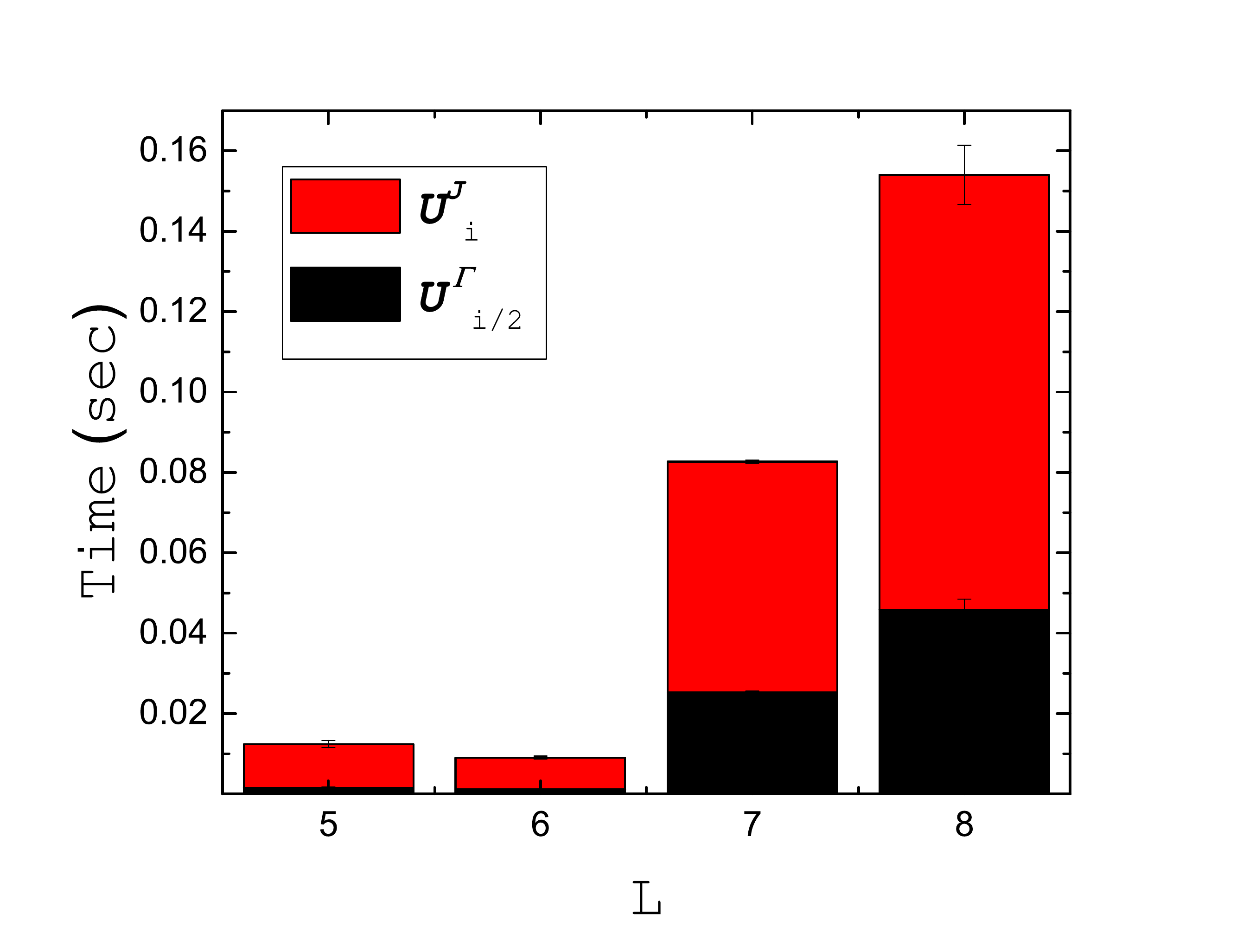}\\
\includegraphics[width=0.6\linewidth]{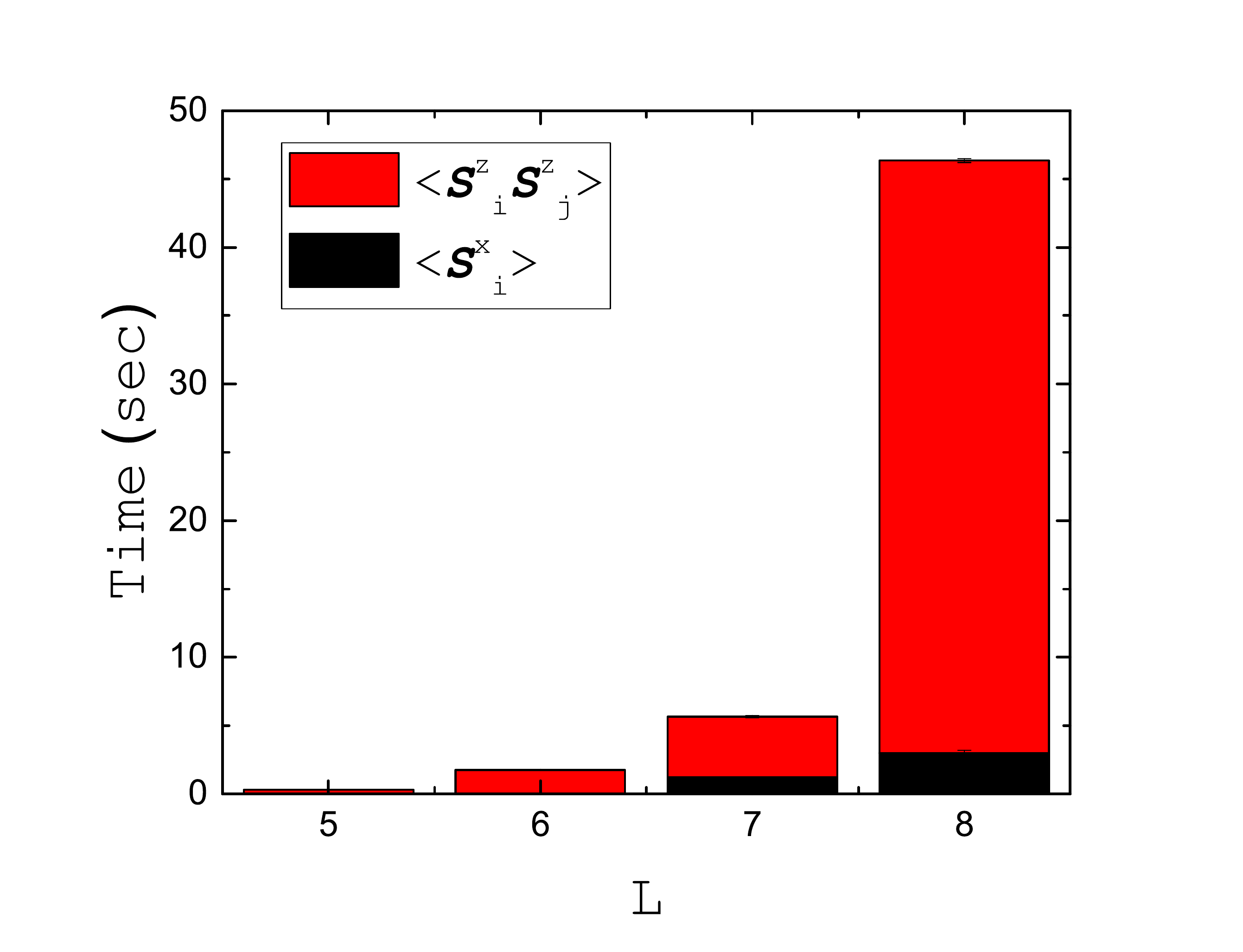}}
\end{center}
\caption{The time to: (top) apply evolution operators,
  $U^{\Gamma}_{\delta\tau/2}$ and $U^J_{\delta\tau}$, and (bottom) to
  calculate expectation values, $m^x_i = \langle \hat{S}^x_i \rangle$
  and $c_{ij} = \langle \hat{S}_i^z \hat{S}_j^z \rangle$, for a
  lattice of size $5 \le L \le 8$ with maximum $\chi =2$ and $\Gamma = J =
  1$. The time to calculate expectation values dominates the execution
  time of the imaginary time evolution algorithm.}
\label{OperatorTime}
\end{figure}

Figure~\ref{OperatorTime} shows that the most computationally
demanding portion of the algorithm is the calculation of expectation
values. For a $L\times L$ lattice, the number of contractions are
$L^2$ and $2L(L-1)$ for the calculation of $m^x_i$ and $c_{ij}$,
respectively. Optimization of this portion of the algorithm is the
chief motivation behind the previously discussed geometry specific
lattice partitioning and quadrant contraction ordering.

\section{Experimental Results}
\label{ExperimentalResults}

The code implementing the ITE algorithm discussed in the previous
section was run on the AWS Elastic Computer Cluster (EC2). We used the
X1.32x large EC2 instance. This instance runs four Intel Xeon E7
8880v3 processors, offering 1952 GiB of DRAM and up to 128 virtual
CPUs (vCPUs). All contractions between pairs of tensors were performed
by folding the tensors into matrices and utilizing the optimized Basic
Linear Algebra Subprogram (BLAS) Double-precision General Matrix
Multiplication (DGEMM) routine \cite{openblas2017}.

\begin{figure}[htb!]
\centering
\includegraphics[scale=0.3]{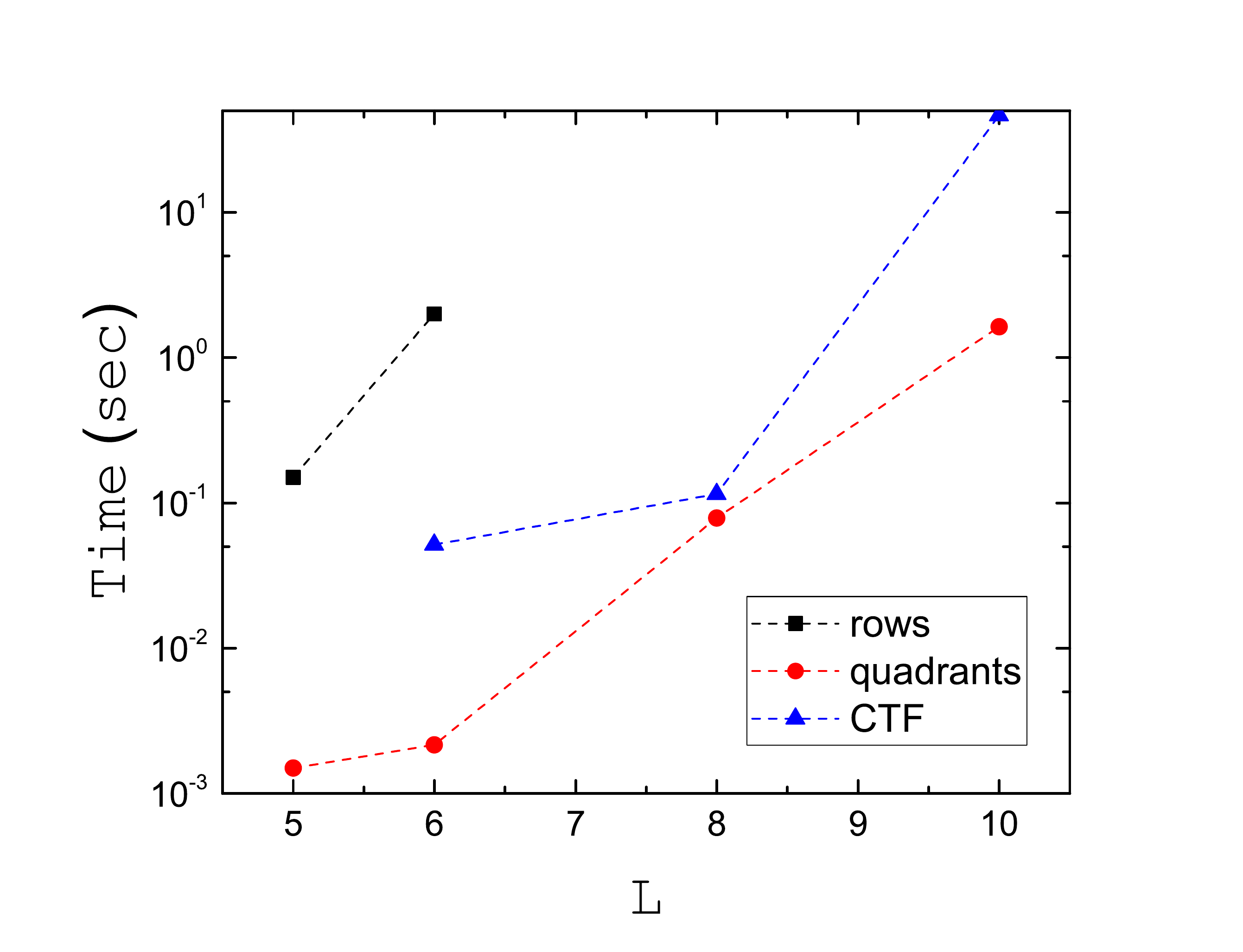}
\caption{A comparison of the time taken to complete a single full
  tensor network contraction for row contraction, quadrant
  contraction, and the CTF contraction algorithms for $L\times L$
  lattice sizes in the range $5 \le L \le 10$ and $\chi = 2$. Larger sizes for the row contraction could not be obtained because of memory constraints.}
\label{ExecutionTimeQCvsRC}
\end{figure}

\begin{figure}[htb!]
\centering
\includegraphics[scale=0.3]{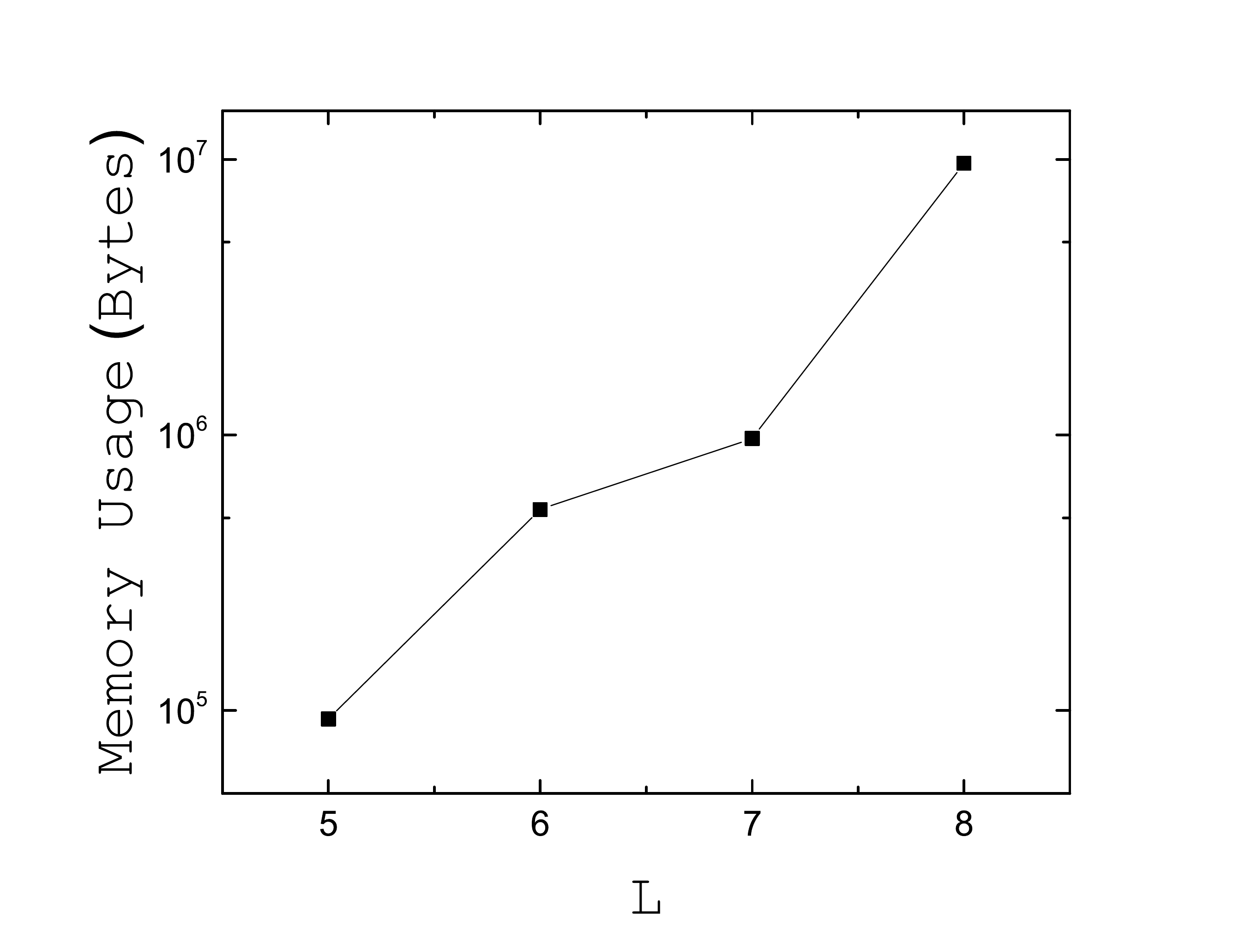}\\
\includegraphics[scale=0.3]{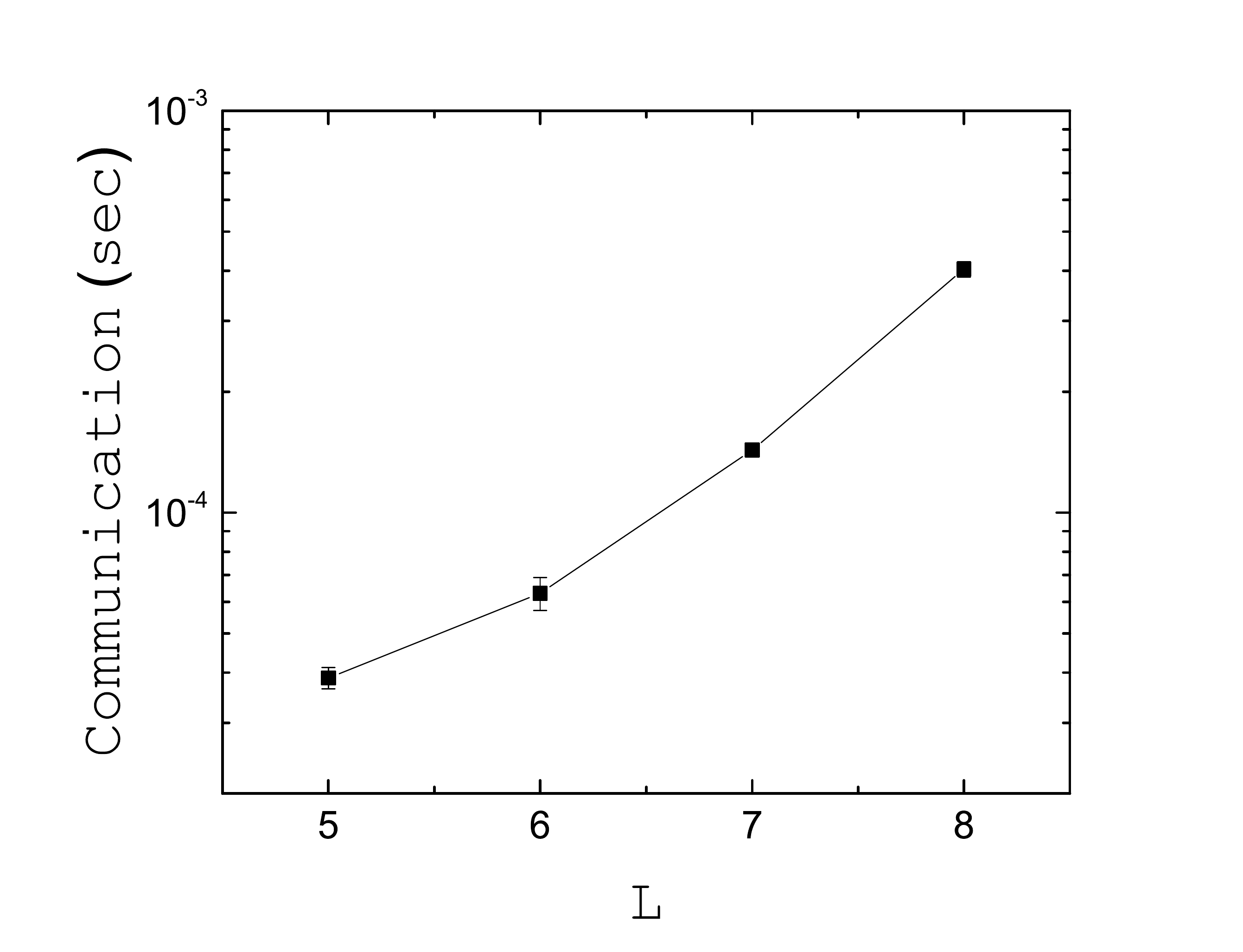}
\caption{(Top) The maximum memory used. (Bottom) The communication
  time during contractions of $L\times L$ lattices of size in the
  range $5 \le L \le 8$ and bond dimension $\chi = 2$.}
\label{MemoryAndCommunicationTime}
\end{figure} 

\begin{figure}[htb!]
\centering
\includegraphics[scale=0.3]{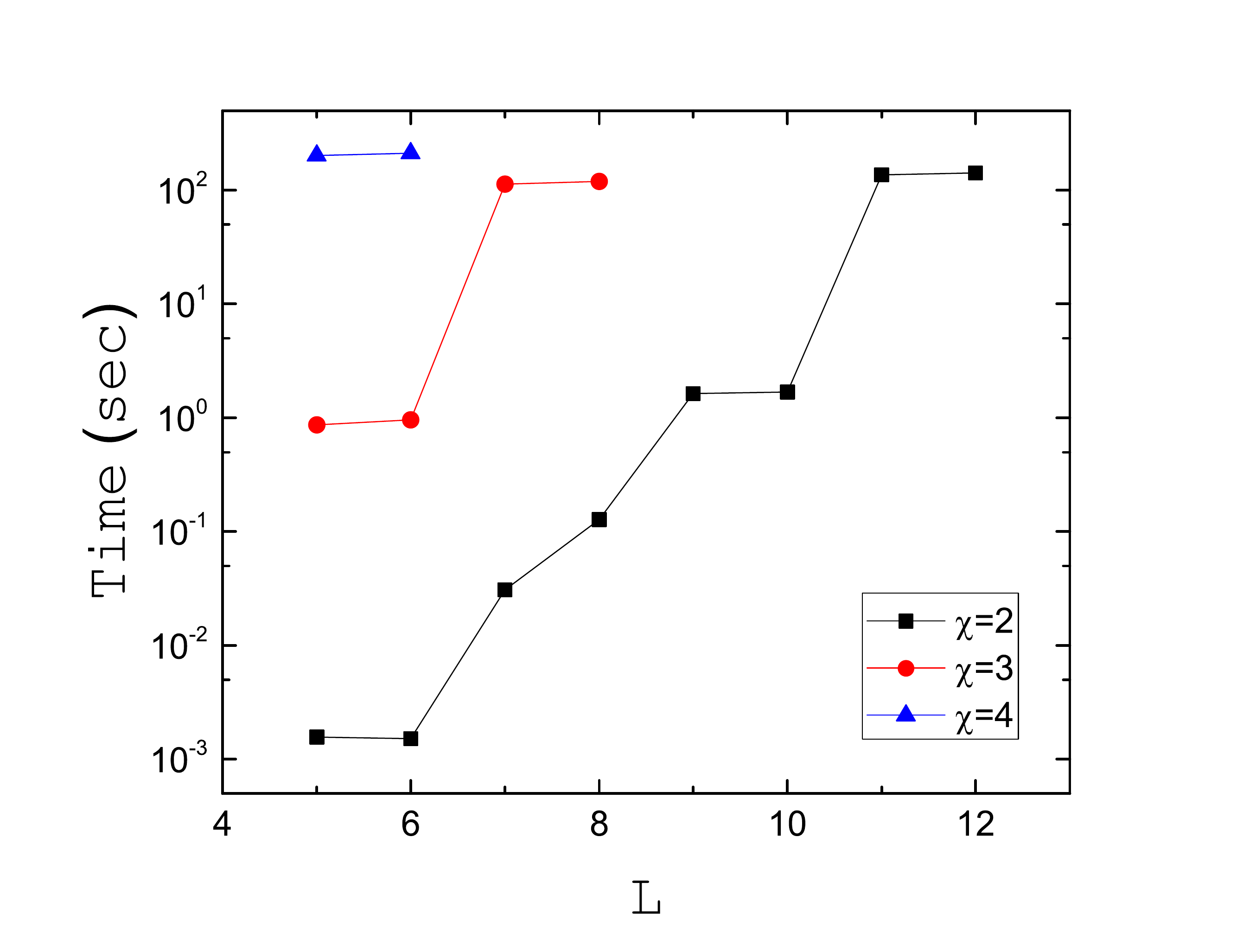}
\caption{Computation time versus uniform bond dimension size $\chi$
  for lattices with length $5 \le L \le 12$. The final point on each line is
  indicative of the largest possible lattice size given bond dimension
  $\chi$. The lattice size is limited by the largest single matrix
  that can be stored in the L3 cache of a multicore processor hosting
  several vCPUs}
\label{ExecutionTimeLargestLattice}
\end{figure} 


\begin{figure}[htb!]
\centering
\includegraphics[scale=0.3]{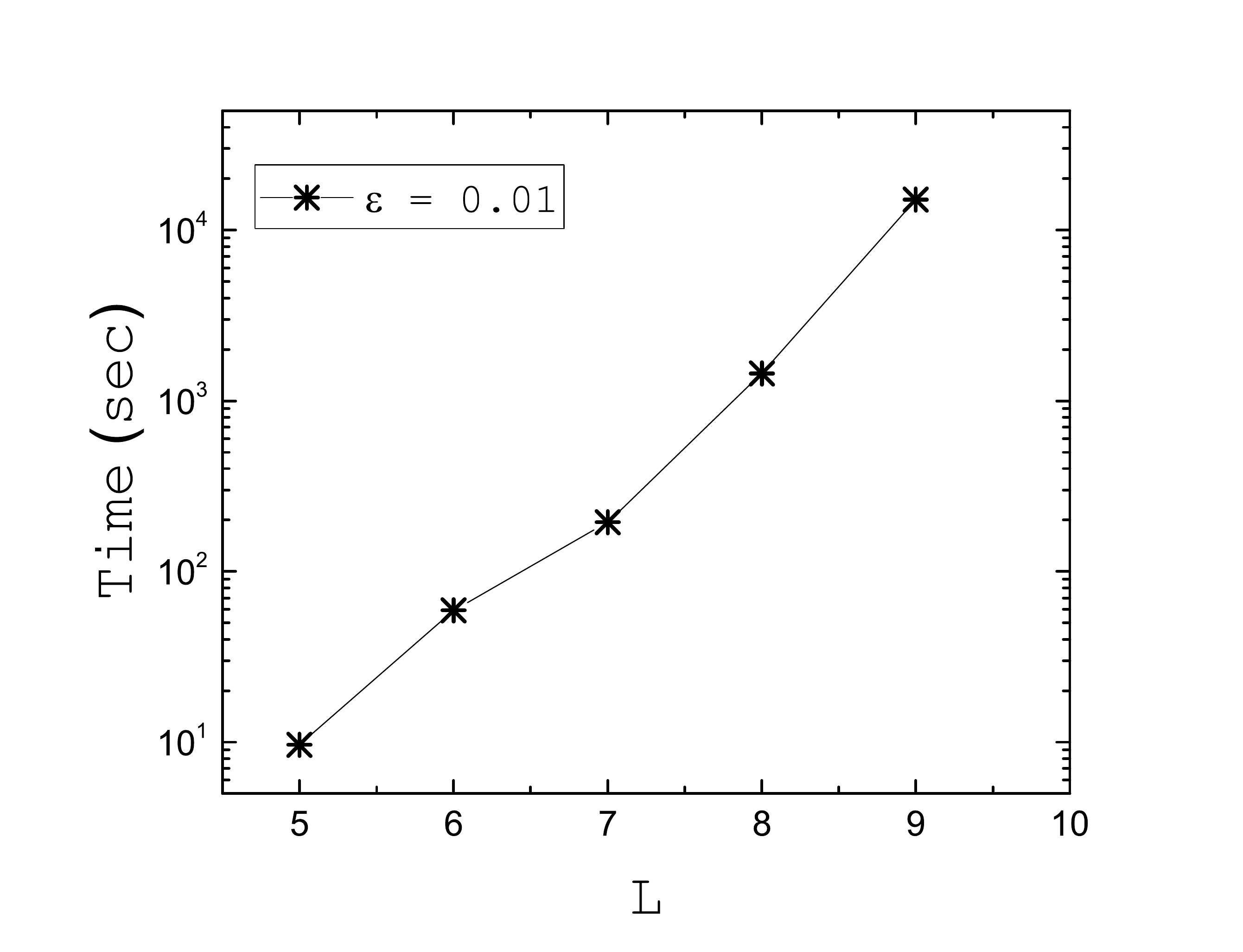}
\caption{The execution time of the ITE as a function of the lattice
  linear size for the Ising model with transverse field strength
  $\Gamma = 1$, at fixed coupling strength $J=1$. The singular value
  cutoff was fixed to $\epsilon = 0.01$, $\tau = 3$, and $n=100$
  steps.}
\label{ExecutionTimeFunctionOfLattice Size}
\end{figure}

\begin{figure}[htb!]
\centering \includegraphics[scale=0.3]{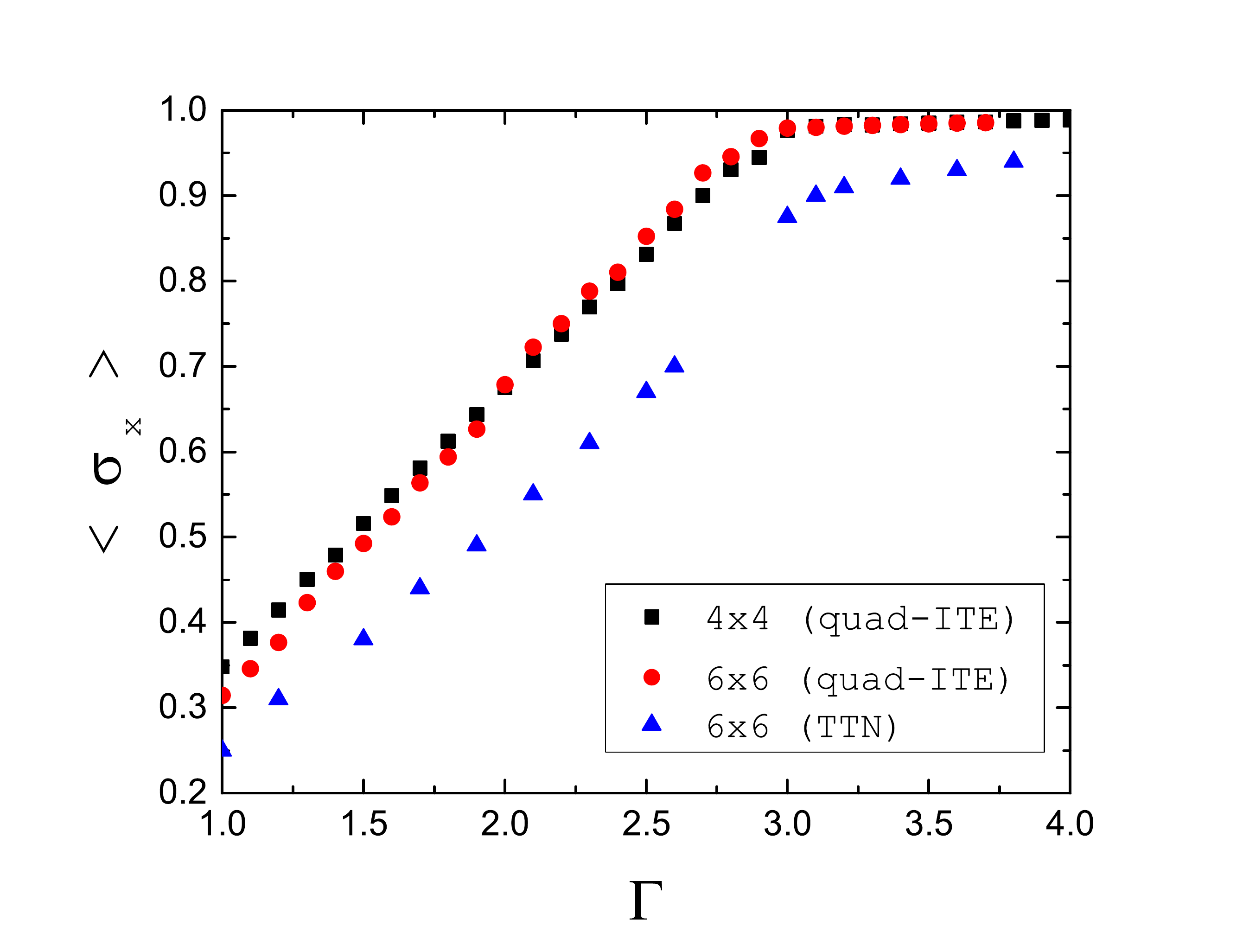}
\caption{The expectation value of the on-site transverse magnetization
  $\langle \sigma_x \rangle$ versus the transverse field strength
  $\Gamma$ for $J=1$. These results are shown to qualitatively agree
  with tree tensor network (TTN) calculations carried out by
  Tagliacozzo {\it et al}. \cite{Tagliacozzo2009}. An elbow indicative
  of a phase transition is clearly observed near $\Gamma = 3$.}
\label{SigmaXObservable}
\end{figure}

\begin{figure}[htb!]
\centering
\includegraphics[scale=0.3]{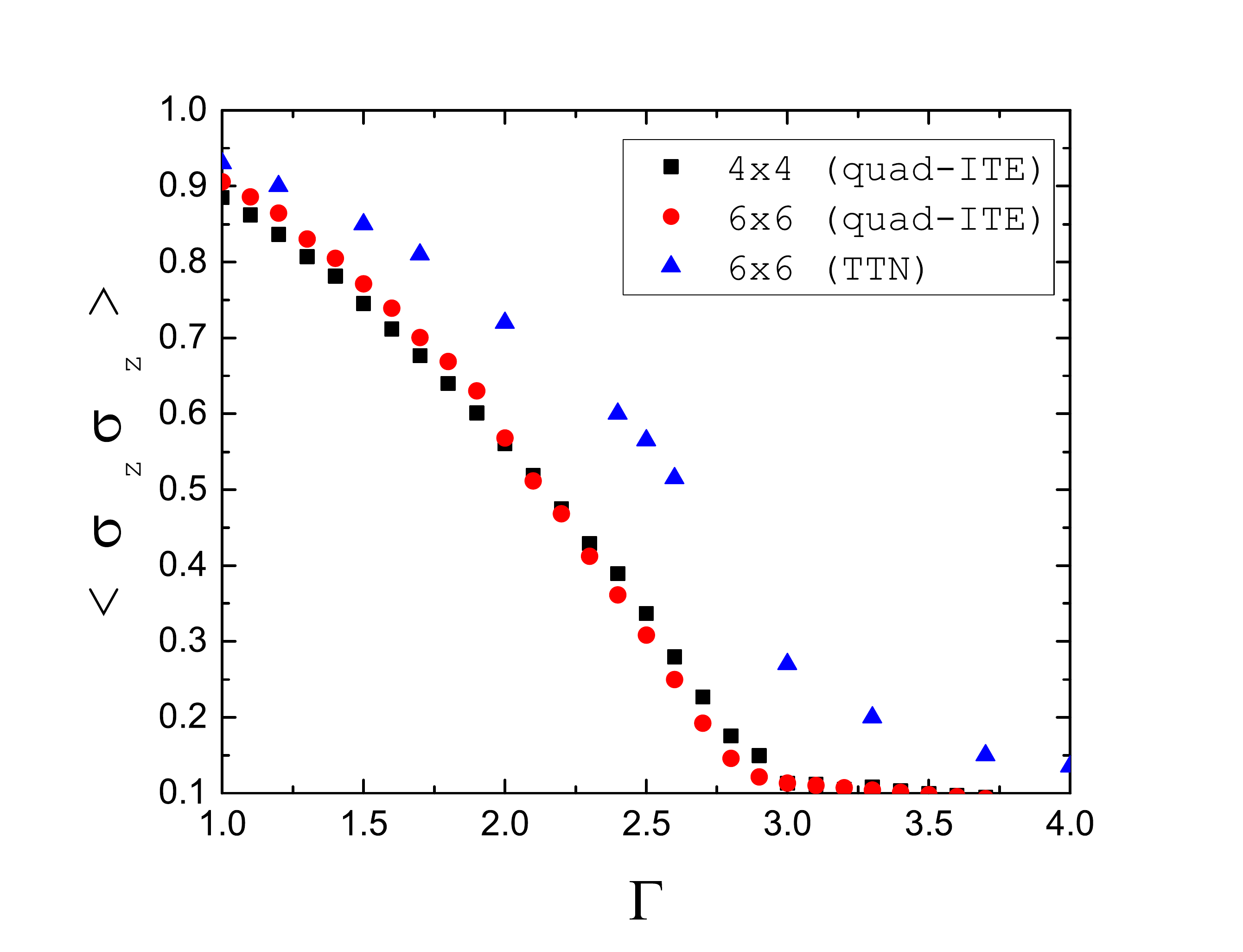}
\caption{The expected value of the longitudinal spin-spin local
  correlator $\langle \sigma_z \sigma_z \rangle$ versus the transverse
  field strength $\Gamma$ for $J=1$. Similarly to
  Fig. \ref{SigmaXObservable}, these results are also qualitatively in
  agreement with those obtained with tree tensor networks
  \cite{Tagliacozzo2009}. The expectation value saturates to a minimal
  value near $\Gamma = 3$, which is indicative of a phase transition.}
\label{SigmaZObservable}
\end{figure}


As a first test of the algorithm performance, we compare each
contraction scheme previously mentioned in order to verify whether an
analysis of the bottleneck tensors is beneficial in determining the
optimal algorithm for tensor contraction. We then determine the limits
of the algorithm and the types of lattices it can handle. We also
compare this to a lattice contracted using the Cyclops Tensor
Framework library (CTF) \cite{Solomonik2012}. This library focuses on
the parallelization of individual tensors by distributing tensor
elements cyclically across processors for generic lattice geometries,
as opposed to our algorithm which is optimized for a specific lattice
geometry. Because of the cyclic distribution of individual tensors across many processors, we avoid using either the previously mentioned row or quadrant contraction orders for the CTF algorithm. Instead, we contract in an order that utilizes $O(D^L +1)$ parameters per contraction.  

Figure ~\ref{ExecutionTimeQCvsRC} shows a comparison of the execution
time is given for the quadrant contraction scheme, the row contraction scheme,
and the CTF contraction for a square lattice size $L \times L$ with every element set to value $1$, and having uniform bond dimension $\chi$ . Quadrant contraction is more efficient than
the row contraction, as anticipated by the analysis of each
algorithm's bottleneck tensors.  For the lattice sizes considered, the
quadrant contraction is also demonstrated to be favorable over the
CTF, proving the importance of lattice-specific optimizations in the
cloud environment. Focusing therefore on the quadrant contraction, we
also determined the average memory and communication costs for various
system sizes. Figure ~\ref{MemoryAndCommunicationTime} shows that the
memory footprint and the communication costs grow exponentially with
the system size.

To analyze the interplay between the bond dimension and the lattice
size, we allowed for larger bond dimensions representing systems in
regimes where entanglement is expected to be large (e.g., near
critical points). Figure ~\ref{ExecutionTimeLargestLattice} shows the
execution time for systems of various sizes, up to a uniform bond
dimension $\chi=4$, again contracting lattices with each element set to value $1$.
 The maximum lattice sizes for $\chi = 2$ and
$\chi=4$ are $L=12$ and $L=6$, respectively, based on the L3 cache of
the AWS instanced used.


We performed the imaginary time evolution simulation of the Ising
model with a transverse field strength of $\Gamma = 1$ and coupling
strength $J = 1$ for varying lattice sizes $L$, measuring the time to
convergence. Time steps were fixed to $\delta\tau = 3/100$. The
singular value cutoff parameter was fixed to $\epsilon = 0.01$, where
$\lambda_k/\lambda_1 \geq \epsilon$, to temper the growth of the bonds
in the system. As anticipated, the time for completion of the ITE
scales exponentially with the system size $L$, as shown in
Figure \ref{ExecutionTimeFunctionOfLattice Size}.

Moving away from systems of low entanglement, we also performed
simulations of the ITE at a transverse field strength of $\Gamma = 3$
and coupling strength $J = 1$ (i.e., near the critical point). The
lattice lengths tested were $L=6$ and $L=8$. For $L=6$ and $L=8$, each
time step $\delta$ was set to $\delta\tau = 3/75$ and $\delta\tau =
4/250$, respectively. The ground states were reached with a
computational runtime of 12.5 hours and 293.3 hours. As anticipated,
bond dimensions grew larger in this regime, reaching its maximal value
at $\chi = 4$.

A final set of experiments for transverse field values $\Gamma$ in the
range $[0,4]$ were run to compare the convergence of the observables
$M^x$ and $M^z$ with previous work established by Tagliacozzo {\it et
  al}. \cite{Tagliacozzo2009} using tree tensor networks (TTNs). Here,
$M_x = (2/N) \sum_i m^x_i$ and $C_{zz} = 1/[N(N-1)] \sum_{\langle i
  j\rangle} c_{ij}$, where $N=L^2$ is the number of lattice sites. The
results are shown in Figures \ref{SigmaXObservable} and
\ref{SigmaZObservable}. The observables we calculate are not in exact
agreement, which we hypothesize to be indicative of the necessity of
larger bond dimensions $\chi$. They do, however, qualitatively exhibit
the behavior indicative of a phase transition near $\Gamma = 3$. At
the paramagnetic phase ($\Gamma > 3J$), the on-site transversed
magnetization $\langle \sigma_x \rangle$ is maximum, while in the
ferromagnetic phase ($\Gamma < 3J$) the on-site longitudinal spin-spin
local correlator $\langle \sigma_z \sigma_z \rangle$ is maximum
instead.
\section{Conclusions}
\label{Conclusions}

At this time supercomputers built around low-latency interconnection
networks are still the best and often the only option for running
scientific and engineering codes. The discrepancy in computational capability between supercomputers and cloud services is obvious when considering tensor network contractions. Distributing tensors over the multiple nodes of a cloud instance increases the communication intensity, therefore slows down the execution on a cloud with relatively high communication latency. The advent of TPUs only further amplifies the gap between computing and communication speed and will shorten the execution time of small size networks; whenever tensors are distributed across multiple nodes only a network with low latency will show significant performance advantage.

The communication latency of cloud services motivated our approach of using large memory instances which would perform large portions of computation with little inter-process communication. Cloud supplies significantly cheaper computing resources than
supercomputers and our results show that this approach can be used for
a range of problems of interest. Several EC2 instances types offer
cost effective alternatives to simulating quantum many-body systems
with tensor networks, such as the \texttt{x1.32x large} instance used in our analysis.

Our tensor network contraction implementation minimizes the
communication costs but is limited by the memory size of the
vCPU. This limitation is determined by the largest single matrix that
can be stored in the L3 cache hosting several vCPUs, a trade off we
make to minimize communication. Addition of L4-level caches advocated fro Big Data applications would also benefit tensor network contraction. 

For the particular example of the two-dimensional Ising model in the presence of a transverse field,
this limitation in memory size appears for lattice sizes of $6\times
6$ (i.e., for a $2^{36}$ dimensional Hilbert space), causing some
deviation from the expected results when the quantum entanglement in
the simulated system is high (i.e. at $\frac{\Gamma}{J} = 3$). This deviation is a result of fixing the maximal bond dimension to $\chi =4$ without introducing any sufficient TN environment approximation scheme (a necessary step to stay within the storage limits available on the cache). Yet, the results are encouraging given the affordability and accessibility
of EC2 instances.

The tensor contraction procedures discussed in Sections~\ref{TNC} and ~\ref{ExperimentalResults} perform better that the Cyclops library for
square lattices with $ L \le 10$. Imminent advancements in the cloud
infrastructure will benefit applications exhibiting fine-grained
parallelism including QMB simulations. Increased memory and larger L1
and L2 caches of individual cores, as well as larger last-level cache
of multicore processors will increase the range of problems that can
be solved without the benefit of tensor partitioning, as described in
this paper. It is unclear if faster networks will make tensor
partitioning more appealing as the communication complexity of very
large problems is likely to increase faster than the benefits due to
lower communication latency.
 
Profiling the code for imaginary time evolution shows that the runtime
of the algorithm is dominated by two in house procedures,
\texttt{contract tVtl par} and \texttt{contract lattice mpi}. The
first, responsible for the contraction of a tensor pair, dominates the
execution time and is called 31,347 times for the $6x6$ ITE run,
accounting for 97.18 \% of the execution time. \texttt{contract
  lattice mpi}, a subroutine responsible for the parallelized
contraction of a lattice, is the parent function of \texttt{contract
  tVtl par} and it is called a total of 3,843 times. Within this
subroutine itself, the time per call is dominated by the calls made to
the \texttt{contract tVtl par} procedure, taking 1.07 ms of the total
1.1 ms per call. This validates the assumption that the algorithm
bottleneck is the contraction of the two largest tensors. Further
confirmation of this is found in the low communication cost, evidenced
by each call to \texttt{contract lattice mpi} requiring 4 calls to
\texttt{MPI Send} and 2 calls to \texttt{MPI Recv}, and yet only
accounting for a combined 2.7 \% of the time per call for
\texttt{contract lattice mpi}.

This analysis shows that the search for optimal contraction algorithms
for tensor networks is critical for solving increasingly large
problems \cite{Pfeifer2014}. An optimal tensor network contraction
algorithm could reduce significantly the communication complexity and
the memory footprint required for the contraction of the largest
tensors, the bottleneck of the process discussed in Sections ~\ref{TNC} and ~\ref{ExperimentalResults}.

In addition, it shows that by properly partitioning a given lattice
(so as to minimize inter-process communication), and selecting
appropriate contraction orderings, cloud services can be effectively
utilized for QMB simulation applications. This is particularly true in
the regime where bond dimensions remain manageable (i.e. away from
critical points). Further work will investigate hybrid algorithms
involving both individual tensor parallelization (as in CTF) and the
geometry specific parallelization approach (as in our algorithm); we
will also explore different lattice geometries under the heuristic of
the minimal bottleneck tensor contraction.

\bibliographystyle{unsrt}
\bibliography{biblio}

\end{document}